\documentclass[%
 reprint,
 superscriptaddress,
showpacs,preprintnumbers,
 amsmath,amssymb,
 aps,
 prl,
 longbibliography,
]{revtex4-1}

\usepackage{graphicx}
\usepackage{dcolumn}
\usepackage{bm}
\usepackage{color}
\usepackage{ulem}

\begin{document}

\preprint{APS/123-QED}

\title{
Dirac half-metal in a triangular ferrimagnet
}

\author{Hiroaki Ishizuka}
\affiliation{
Department of Applied Physics, University of Tokyo, Hongo, 7-3-1, Bunkyo, Tokyo 113-8656, Japan
}

\author{Yukitoshi Motome}%
\affiliation{
Department of Applied Physics, University of Tokyo, Hongo, 7-3-1, Bunkyo, Tokyo 113-8656, Japan
}

\date{\today}

\begin{abstract}
An idea is proposed for realizing a fully spin-polarized Dirac semimetal in frustrated itinerant magnets.
We show that itinerant electrons on a triangular lattice exhibit the Dirac cone dispersion with
half-metallic behavior in the presence of a three-sublattice ferrimagnetic order.
The Dirac nodes have the same structure as those of graphene.
By variational calculation and Monte Carlo simulation, we demonstrate that the ferrimagnetic order with
the Dirac node spontaneously emerges in a simple Kondo lattice model with Ising anisotropy.
The realization will be beneficial for spintronics as a candidate for spin-current generator. 
\end{abstract}

\pacs{
71.10.Fd, 73.22.Pr, 72.25.-b
}
\maketitle

Massless Dirac fermions show substantially different nature from ordinary electrons. 
The peculiar nature originates in the characteristic energy dispersion 
---the nodal structure with linear dispersion often referred to as the Dirac cone.
While the Dirac fermions were originally introduced in the relativistic quantum theory,
recent discovery of graphene~\cite{Novoselov2004,Novoselov2005}, a single layer sheet of graphite,
has carved out a new direction of their study in condensed matter systems~\cite{Geim2007,CastroNeto2009}.
In graphene, two Dirac cones appear in the energy dispersion of $\pi$ electrons, which are
at the $K$ and $K^\prime$ points in the Brillouin zone for the two-dimensional honeycomb lattice.
The Fermi level of this two dimensional conductor 
comes right at the nodal points, and the low-energy
Hamiltonian is well approximated by the Weyl equation~\cite{Semenoff1984}.
Various remarkable electronic and transport properties of graphene mainly owe to these Dirac cones in
the band structure.

The extraordinary nature of Dirac fermions in graphene has also attracted a great interest from
application to electronics~\cite{Geim2007}.
From the viewpoint of such potential applications, it is of great interest to control the characteristic
band structure. 
Furthermore, it is also desired to control the electronic spin degree of freedom for the application to
spintronics~\cite{Wolf2001}.
However, there is not so much flexibility in graphene, as the Dirac cone is a direct consequence of the
honeycomb lattice geometry and the relativistic spin-orbit interaction is very weak.

In this Letter, we propose an alternative solution for manipulating the spin degree of freedom 
by seeking possible emergence of Dirac fermions from itinerant magnets.
We show that itinerant electrons coupled to a well-known ferrimagnet on a triangular lattice give rise to the
Dirac nodes in their band structure, similar to those of graphene. 
The resultant massless Dirac fermions are spin-polarized, and they are stable in a wide range of the spin-charge
coupling including typical values in solids.
We demonstrate that, by an unbiased Monte Carlo (MC) simulation as well as a variational calculation, such Dirac half-metal with ferrimagnetic order
spontaneously emerges in a minimal Kondo-lattice type model.
The results strongly suggest the possibility of realizing the exotic electronic state in transition-metal and
rare-earth compounds, which generally retain much higher controllable degrees of freedom than graphene.
Such a new family will not only add a member to the known list of Dirac electrons in
solids~\cite{Cohen1960,Konoike2012,Hasan2010,Ishibashi2008}, but also bring a completely new aspect by the spin
polarization.
In a half-metal, the electric current is perfectly spin-polarized as the low energy excitations only exist for 
the majority spin~\cite{Wolf2001}. This nature works as a spin-current generator by filtering out the minority-spin electrons.
Thus, our proposal opens a new frontier for the application of Dirac massless fermions, especially for
spintronics~\cite{Pesin2012}.

\begin{figure}
   \begin{center}
   \includegraphics[width=0.7\linewidth]{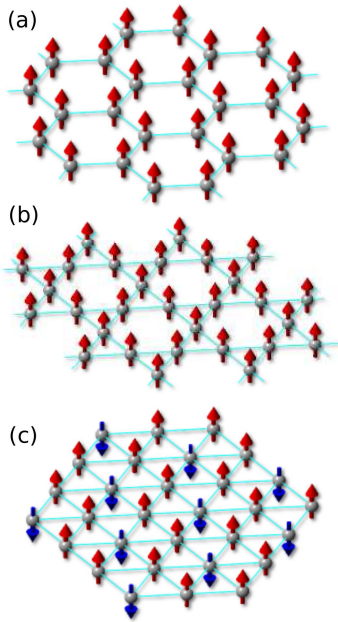} 
   \caption{
   (color online). Schematic pictures of (a) a honeycomb ferromagnet, (b) kagome ferromagnet, and
   (c) three-sublattice triangular ferrimagnet. 
   The arrows at each site represent localized spins. 
   }
   \label{fig:model}
   \end{center}
\end{figure}

\begin{figure*}
   \begin{center}
   \includegraphics[width=0.9\linewidth]{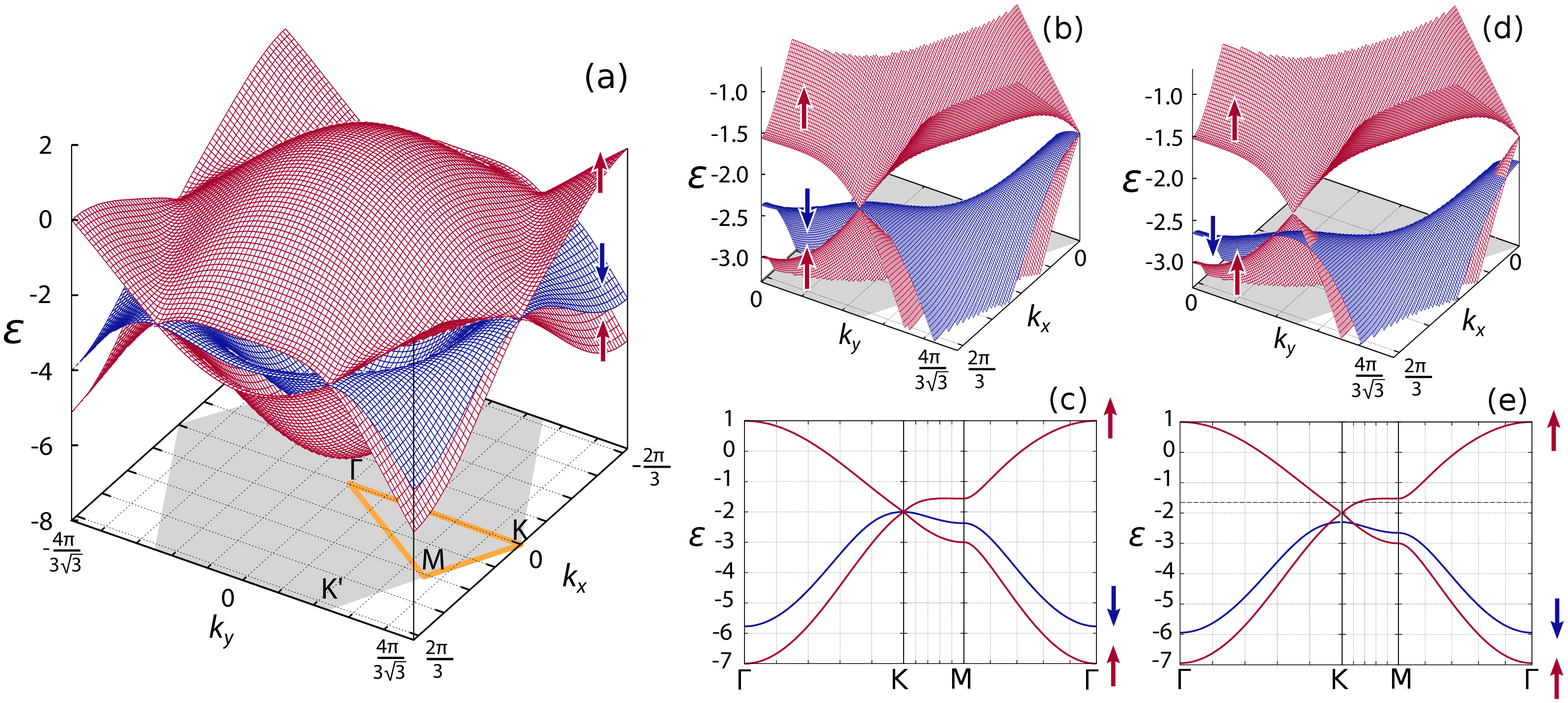} 
   \caption{
   (color online). Band structures of the model in Eq.~(\ref{eq:H_DE}) under the three-sublattice ferrimagnetic
   order at $J=2$.
   (a) The overall band structure of the three lower-energy bands at $J^\prime=0$, (b) the enlarged view near
   the Fermi level $\varepsilon=-J$ at $n=1/3$ in the first quadrant, and (c) the cut along the symmetric lines.
   (d) and (e) show the results at $J^\prime =0.05$.
   The arrows indicate the spins for each band. 
   In (a), the gray hexagon on the basal plane shows the first Brillouin zone for the magnetic supercell.
   The dashed line in (e) indicates the Fermi level in the MC simulation shown in Fig.~\ref{fig:mcdata}.
   }
   \label{fig:band}
   \end{center}
\end{figure*}

Let us first discuss a naive, rather trivial approach to achieve a Dirac half-metal. 
We here consider a single-band ferromagnetic Kondo lattice model (double-exchange model) on a honeycomb or
kagome lattice [see Figs.~\ref{fig:model}(a) and \ref{fig:model}(b)].
The model consists of the nearest-neighbor hopping of electrons and the exchange interaction between the
electron spin and localized moment, whose Hamiltonian is given by 
\begin{eqnarray}
H = -t \! \sum_{\langle i,j \rangle, \sigma} \! ( c^\dagger_{i\sigma} c_{j\sigma} + \text{H.c.} ) - J \sum_{i} {\bm \sigma}_i \cdot \mathbf{S}_i.
\label{eq:H_DE}
\end{eqnarray}
Here, $c_{i\sigma}$($c_{i\sigma}^\dagger$) is the annihilation(creation) operator of an itinerant electron with
spin $\sigma=\uparrow, \downarrow$ at $i$th site, $\bm \sigma_i$ and $\mathbf{S}_i$ represent the itinerant and
localized spin, respectively; $t$ is the transfer integral and $J$ is the onsite Kondo coupling.
Hereafter we take $t=1$ and $J>0$.

In this model, when $J$ is sufficiently large compared to the bandwidth at $J=0$, a ferromagnetic order is
stabilized by the double-exchange mechanism in a wide range of electron filling
$n=\sum_{i\sigma} \langle c_{i\sigma}^\dagger c_{i\sigma} \rangle / 2N$, where $N$ is the system
size~\cite{Zener1951,Anderson1955}. 
In the ferromagnetic phase, the band is split into two by the large exchange coupling according to
the spin component, and each band has exactly the same form as that for the noninteracting case $J=0$.
Hence, in principle, the Dirac half-metal arises for the honeycomb and kagome lattices, as the noninteracting
bands on these lattices have the Dirac nodes.
However, these situations are very difficult to realize in solids as neither such a strong exchange interaction
nor the honeycomb and kagome structures is easily realized in magnetic compounds.

As a more realistic approach, here we propose a simple, but rather nontrivial route to the half-metallic Dirac
fermion systems.
Let us consider the model in Eq.~(\ref{eq:H_DE}) on a triangular lattice, and the situation in which a
three-sublattice collinear ferrimagnetic order with up-up-down spin configuration is realized
---see Fig.~\ref{fig:model}(c).
By treating the localized moments as classical spins with $|\mathbf{S}_i| = 1$, the band
structure is easily calculated by the exact diagonalization of the Hamiltonian.
The lower three bands of the totally six bands are shown in Fig.~\ref{fig:band}; the two red bands are
of up spins, and the blue band is of down spin (the other upper three bands have the similar form with
opposite spins).

The band structure has a notable feature at the energy $\varepsilon = -J$; the two up-spin bands touch with
each other at the $K$ and $K^\prime$ points in the Brillouin zone to form a Dirac-type point node with linear
dispersion, and the down-spin band has the band top at the same points with an ordinary parabolic dispersion. 
See also the enlarged figure in Fig.~\ref{fig:band}(b) and the energy dispersion along the symmetric lines in
Fig.~\ref{fig:band}(c).

In this situation, when the electron filling is at $n=1/3$, the two lower bands are fully occupied while the
remaining bands (including the upper three) are unoccupied; the Fermi level is located at the nodes where the
three bands meet.
As the down-spin band has an energy gap, the half-metallic Dirac electrons are obtained by electron doping to
the unoccupied up-spin band.
Although hole doping hides the Dirac nature as the down-spin parabolic band is doped at the same time,
the situation is avoided by introducing an additional antiferromagnetic exchange coupling between the
neighboring sites, $J' \sum_{\langle i,j \rangle} {\bm \sigma}_i \cdot \mathbf{S}_j$~\cite{note_jprime}. 
A finite $J'>0$ shifts the down-spin band to the lower energy and isolates the half-metallic Dirac nodes
energetically, as demonstrated in Figs.~\ref{fig:band}(d) and \ref{fig:band}(e).
Hence, the simple ferrimagnetic order on the triangular lattice realizes the peculiar Dirac half-metallic state
near 1/3 filling. 

The Dirac nodes have essentially the same structures as those in graphene. 
Under the ferrimagnetic order, the Hamiltonian is written as 
\begin{eqnarray}
{\cal H} = 
\sum_{\bf k}
\begin{pmatrix}
-J \sigma^z_{A} & \tau_{\bf k} & \tau_{\bf k}^\ast \\
 \tau_{\bf k}^\ast & -J \sigma^z_{B} & \tau_{\bf k} \\
 \tau_{\bf k} & \tau_{\bf k}^\ast & \left( J + 6J^\prime \right) \sigma^z_{C}
\end{pmatrix}
.
\label{eq:ferri}
\end{eqnarray}
Here, the upper two rows correspond to the sites with the up localized moment ($A$, $B$ sublattices) and the bottom row is for the down one ($C$ sublattice) in the three-site unit cell.
In Eq.~(\ref{eq:ferri}), $\sigma^{z}$ is the $z$ component of the Pauli matrix for itinerant electrons,
$\bf k$ is the wave vector, and $\tau_{\bf k}$ is the Fourier transform 
of the hopping term given by
$
\tau_{\bf k} = -t[e^{{\rm i}k_x} + e^{{\rm i}\left(-\frac{k_x}2 + \frac{\sqrt3}2 k_y\right)} + e^{{\rm i}\left(-\frac{k_x}2 - \frac{\sqrt3}2 k_y\right)}]
$.
By using the $\bf k\cdot p$ perturbation around the $K$ and $K^\prime$ points in the Brillouin
zone~\cite{Semenoff1984} and by expanding the result up to the first order in terms of
$t\kappa_x/J$ and $t\kappa_y/J$ ($\boldsymbol{\kappa}$ is the relative wave vector measured from $K$ and
$K^\prime$ points), we end up with the low-energy Hamiltonian which is factorized into two parts. 
One is a $2\times 2$ Hamiltonian for the up-spin honeycomb subnetwork of the $A$ and $B$ sublattices, 
and the other is a localized state at the down-spin sites in the $C$ sublattice. 
The former is given by 
\begin{eqnarray}
{\cal H}_{{\bf k}\pm}^\text{Dirac} =
\begin{pmatrix}
-J & \frac32{\rm i} t \left( \kappa_x \pm {\rm i}\kappa_y \right)  \\
-\frac32 {\rm i} t \left( \kappa_x \mp {\rm i}\kappa_y \right) & -J
\end{pmatrix}
,
\label{eq:kp}
\end{eqnarray}
where the sign $\pm$ corresponds to the $K$ and $K^\prime$ points.
This has an equivalent form to that of graphene.

It is worthy to note that the Dirac nodes are formed immediately by switching on $J$.
However, when $J$ is small,
the low-energy physics at $n=1/3$ is not characterized solely by the massless Dirac fermions 
because there is a band overlap at the energy of the Dirac nodes. 
The band overlap comes from the second lower band for up spin, which has an energy minimum at
${\bf k}=(2\pi/3,0)$ points and its threefold symmetric points for small $J$;
the minimum energy is given by $\varepsilon_{(\frac{2\pi}3,0)}=t/2 - \sqrt{\left(J+3J^\prime-t/2\right)^2+2t^2}$.
In order for the Dirac nodes to be isolated at the Fermi level, this energy should be higher than that
at the $K$ and $K^\prime$ points, $\varepsilon_K=-(J+3J^\prime)$.
Hence, the Dirac nodes are energetically isolated 
and play a decisive role when the condition $(J + 3J^\prime)/t > 1$ is satisfied. 
This condition is important because the necessary $J$ and $J^\prime$ are much smaller than the noninteracting
bandwidth $9t$, and it is indeed satisfied in wide range of materials. 

\begin{figure}
   \begin{center}
   \includegraphics[width=0.7\linewidth]{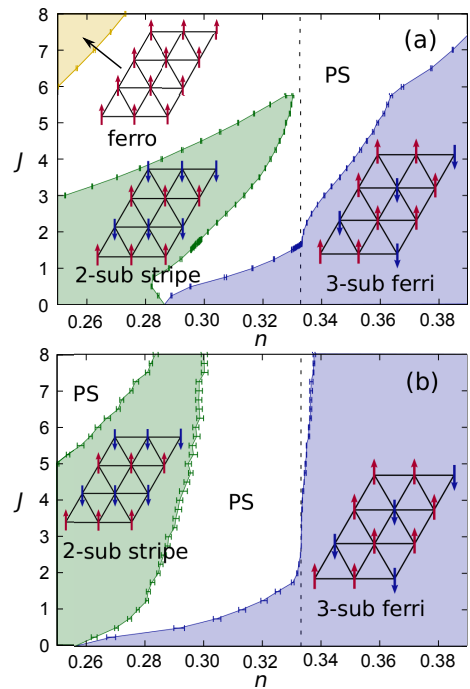} 
   \caption{
   (color online). Ground state phase diagram obtained by variational calculation at (a) $J^\prime = 0$ and
   (b) $J^\prime=0.05$.  The schematic picture of magnetic structure in each phase is shown.
   The white region indicates the electronic phase separation (PS) and the dotted vertical lines indicate
   $n=1/3$.
   }
   \label{fig:diagram}
   \end{center}
\end{figure}

So far, we assumed the presence of three-sublattice ferrimagnetic order. 
In the following, we show that such order is indeed stable in the Kondo-lattice type model as
Eq.~(\ref{eq:H_DE}). 
We here simplify the model by assuming the localized moments are the Ising spins taking the values $S_i = \pm 1$.

First, we investigate the ground state phase diagram near $n=1/3$ by a variational calculation.
We compare the ground state energy of the two-sublattice stripe phase and three-sublattice ferrimagnetic phase
appeared in the previous study~\cite{Ishizuka2012}, in addition to the ferromagnetic phase.
The results at $J=2$ are shown in Fig.~\ref{fig:diagram} for $J^\prime = 0$ and $0.05$.
At $J^\prime = 0$, the ground state in the plotted range is dominated by 
the ferrimagnetic phase as well as the stripe phase. 
The different phases are separated by phase separation.
As shown in Fig.~\ref{fig:diagram}(b), the introduction of small $J^\prime$ largely stabilizes
the ferrimagnetic phase near $n=1/3$ as well as the stripe phase. 
This is because the itinerant electron spins are polarized parallel to the localized spins in the ground state, 
leading to an energy gain (loss) by the antiferromagnetic $J^\prime$ for the two states (the ferromagnetic state).

\begin{figure}
   \begin{center}
   \includegraphics[width=0.7\linewidth]{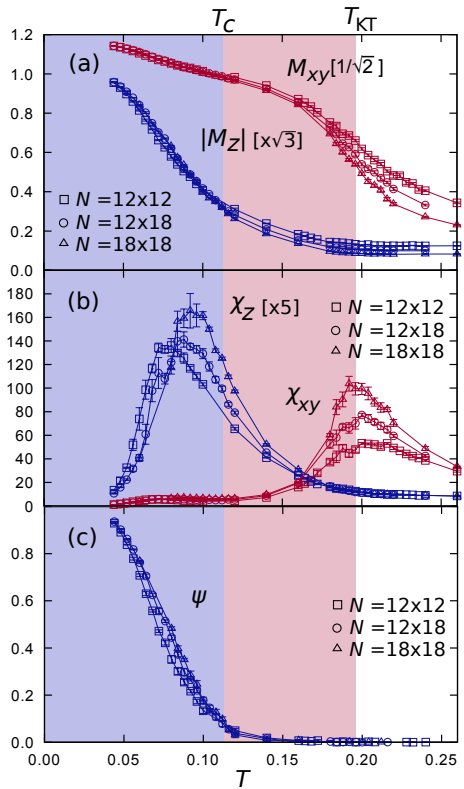}
   \caption{
   (color online). MC results for (a) the pseudo moments $M_{xy}$ and $|M_z|$, (b) corresponding
   susceptibilities $\chi_{xy}$ and $\chi_z$, and (c) azimuth parameter $\psi$. 
   The data are calculated at $n=0.34$.
   }
   \label{fig:mcdata}
   \end{center}
\end{figure}

We next examine the stability of the ferrimagnetic order at finite temperatures by an unbiased MC simulation.
For the simulation, a standard algorithm for fermion systems coupled to classical fields is used~\cite{Yunoki1998}.
In this method, the trace over the fermions in the partition function is calculated by the exact
diagonalization, while the trace over classical spin configurations is computed by a classical MC 
method using the Metropolis dynamics. 
The phase transition to ferrimagnetic phase is detected by using two parameter~\cite{note_pmoment}. 
One is the pseudo-moment defined by 
\begin{eqnarray}
\tilde{\bf S}_m = 
\left(
\begin{array}{ccc}
\frac2{\sqrt6} & -\frac1{\sqrt6} & -\frac1{\sqrt6} \\
0              &  \frac1{\sqrt2} & -\frac1{\sqrt2} \\
\frac1{\sqrt3} &  \frac1{\sqrt3} &  \frac1{\sqrt3} \\
\end{array}
\right)
\left(
\begin{array}{c}
S_i  \\
S_j  \\
S_k  \\
\end{array}
\right),
\end{eqnarray}
where $m$ is the index for the three-site unit cells, and 
$(i,j,k)$ denote the three sites in the $m$th unit cell belonging to the sublattices $(A,B,C)$, respectively. 
We measure the summation ${\bf M} = (3/N) \sum_m \tilde{\bf S}_m$ and the susceptibility. 
The other is the azimuth parameter $\psi$ defined by $\psi = (\tilde{M}_{xy})^3 \cos{6 \phi_M}$, 
where $\phi_M$ is the azimuth angle of $\bf M$ in the $xy$ plane and 
$\tilde{M}_{xy} = 3 M_{xy}^2/8$ ($M_{xy}^2 = M_x^2 + M_y^2$). 
The ferrimagnetic ordering is signaled by $M_{xy} \to 2\sqrt{2/3}$, $|M_z| \to 1/\sqrt{3}$, and $\psi \to 1$
at low temperature $T \to 0$, respectively~\cite{Ishizuka2012,Takayama1983,Fujiki1986}. 

Figure~\ref{fig:mcdata} shows the MC results at $J=2$ and $J^\prime = 0.05$
in the slightly electron doped region to $n=1/3$ [see also Fig.~\ref{fig:band}(e)].
The results indicate two successive phase transitions at $T_{\rm KT}=0.192(15)$ and at $T_c=0.108(9)$. 
The transition temperatures are estimated by extrapolating the peak of susceptibilities $\chi_{xy}$ and
$\chi_z$ as $N \to \infty$.
The transition at $T_{\rm KT}$ is considered as a Kosterlitz-Thouless type with the growth of quasi-long-range
order~\cite{Ishizuka2012}. 
On the other hand, the phase transition at $T_c$ is a three-sublattice ferrimagnetic ordering.
The MC result and the above analysis for the ground state consistently indicate that the
three-sublattice ferrimagnetic order is stabilized in the vicinity of $n =1/3$ in the wide range of
parameters for $J$ and $J^\prime$, spontaneously giving rise to the Dirac half-metal.

As such ferrimagnetic order was indeed observed in several insulating magnets~\cite{Tanaka1989,Iida1993}, 
our results in the minimal model will stimulate the hunt for Dirac half-metal in transition-metal and rare-earth
compounds. 
The present results will be qualitatively robust even when extending the model to more realistic
situation. 
For instance, the ferrimagnetic state remains stable when including the transverse components of localized
spins, at least, in the presence of the Ising anisotropy. 
Multi-band effect may be avoided under a particular crystal field; for instance, the $d$-electron $a_{1g}$
orbital isolated by a strong trigonal field is a good candidate for the realization. 
Interlayer coupling, however, may open a gap at the Dirac nodes.
Nevertheless, a straightforward stacking of layers or sufficiently isolated layers in a controlled thin film
will be promising to preserve the massless nature.

The authors thank Y. Matsushita, A. Shitade, and Y. Yamaji for helpful comments.
H.I. is supported by Grant-in-Aid for JSPS Fellows.
This research was supported by KAKENHI (No.19052008, 21340090, 22540372, and 24340076),
Global COE Program ``the Physical Sciences Frontier", the Strategic Programs
for Innovative Research (SPIRE), MEXT, and the Computational Materials Science
Initiative (CMSI), Japan.


\begin{thebibliography}{99}
\bibitem{note_jprime}    An off-site Kondo coupling may exist generally in the Kondo lattice systems, 
                         although the magnitude is much smaller than the onsite one and 
                         the sign depends on the orbital nature of itinerant and localized electrons. 
                         The antiferromagnetic superexchange coupling between neighboring localized spins, 
                         given by $J_{\rm AF} \sum_{\langle i,j \rangle} \mathbf{S}_i \cdot \mathbf{S}_j$, may also exist, 
                         but it neither modifies the band structure nor harms the stability of the ferrimagnetic state.
\end{thebibliography}
\end{document}